\documentclass[twocolumn,letterpaper]{jpsj2} 
\def\simge{\lower0.7ex\hbox{$\ \overset{>}{\sim}\ $}}
\def\simle{\lower0.7ex\hbox{$\ \overset{<}{\sim}\ $}}

\title{Enhanced Charge Fluctuations Due to Competitions between
Intersite and Kondo-Yosida Singlet
Formations in Heavy-fermion Systems}
\author{Kazumasa \textsc{HATTORI}$^{1}$ and Kazumasa \textsc{MIYAKE}$^2$}

\inst{$^1$Institute for Solid State Physics, University of Tokyo,
Kashiwa, Chiba 277-8581,Japan \\
$^2$Division of Materials Physics, Department of Materials Engineering Science, Graduate School of Engineering Science, Osaka University, Toyonaka, Osaka 560-8531, Japan
}

\abst{
 We investigate f-electron charge susceptibility in a two-impurity
 Anderson model 
on the basis of Wilson's numerical renormalization group method. 
The f-electron charge susceptibility diverges logarithmically at
 the critical point of this model when conduction-electron bands exhibit 
 particle-hole symmetry. Although the 
 critical point disappears without the particle-hole symmetry, 
the f-electron charge fluctuation is
 much more enhanced near the crossover regime between the Kondo-Yosida singlet
  and intersite spin-singlet states than that in the single-impurity
 case. This result shows that charge fluctuations are enhanced owing
 to the competition between intersite and Kondo-Yosida spin singlets. 
 A possible scenario for the enhanced residual 
resistivity near the region where the Kondo temperature becomes
 comparable with the N\'eel temperatures 
 under pressure in some heavy-fermion compounds is proposed.}

\kword{two-impurity Anderson model, numerical renormalization group,
residual resistivity, quantum critical point\vspace{-2cm}}

\begin{document}
\maketitle

 Since the first heavy-fermion compound CeAl$_{3}$ was discovered in
 1975\cite{CeAl3}, many  heavy-fermion f-electron systems have been discovered  and 
investigated.\cite{Review} Many Ce-based compounds 
show a quantum 
critical point (QCP) of its long range ordered phase under various
external conditions such as pressure and magnetic fields. 
Near the QCP, anomalous properties appear in many physical 
quantities such as the diverging
 specific heat coefficient, the non-Fermi liquid (NFL)  
temperature dependence of the resistivity, and  
the enhancement in the residual resistivity. 
Furthermore, an unconventional superconductivity appears 
near the QCP in many compounds\cite{Review}. Among 
them, various anomalous properties
near the QCP associated with 
an antiferromagnetic (AFM) long-range order have attracted much attention. 
A qualitative explanation of QCPs was proposed by Doniach,\cite{Doniach} which is
competition between the intersite antiferromagnetic order and
local Kondo-Yosida singlet formation.

``Conventional'' QCPs  are well described by the 
Moriya-Hertz-Millis (MHM)-type critical theory of itinerant 
magnetism,\cite{Moriya, Hertz, Millis} 
 assuming the existence of quasiparticles and Landau's Fermi liquid 
theory\cite{Landau}.
This means that Kondo-Yosida singlet formation starts even inside the 
AFM phase,  and that f-electrons form strongly dressed 
quasiparticles with a large effective mass in many heavy-fermion 
compounds. This qualitative picture is the basis of the MHM 
theory of itinerant AFM fluctuations in heavy-fermion systems.

In some cases such as CeCu$_{6-x}$Au$_x$\cite{CeCuAuX} with $x=0.1$ and YbRh$_2$Si$_2$\cite{YbRhSi}, 
the conventional MHM-type critical spin fluctuation theory is not 
directly applicable, and thus alternative schemes have been
proposed, in which changes in the Fermi surface are 
emphasized\cite{Alternative1,Alternative2}. In these theories, 
 no Kondo-Yosida singlet is formed inside the AFM phase. 
 In order to explain the anomalous properties observed in these
 compounds, Moriya's spin fluctuation theory is also 
extended\cite{Alternative3}.
In spite of these theoretical studies, the anomalous properties of these
materials remain to  be elucidated.

In this Letter, 
we will examine the effects of intersite AFM  
interactions on the formation of heavy quasiparticles and/or 
 Kondo-Yosida singlets in heavy-fermion systems and show that charge
fluctuations are strongly enhanced by the formation of Kondo-Yosida
singlet states in a two-impurity Anderson model (2IAM).

\begin{figure}[b]
\begin{center}
\includegraphics[scale =0.4]{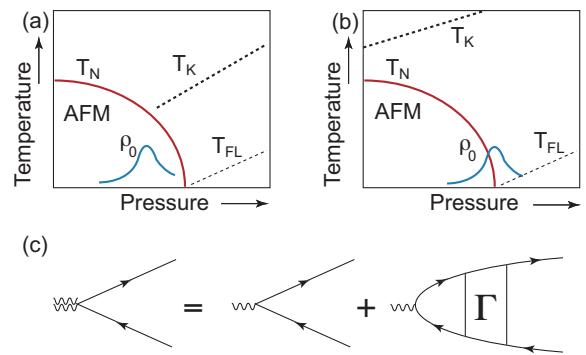}
\caption{(Color online) Schematic temperature-pressure phase diagram of
 Ce-based 
heavy-fermion systems in the case that 
 $\rho_0$ has a peak (a) inside the AFM phase and (b) at
 the QCP. $T_{\rm FL}$ is the so-called Fermi liquid temperature. For Yb-based compounds, the horizontal axis should be regarded
 as a ``negative'' pressure. (c) Diagrammatic representation of effective potential
 scattering (double wavy line). The arrows indicate the one-particle Green's
 function, and the wavy line represents the bare potential scattering. 
$\Gamma$ is the four-point vertex function.}
\label{fig-1}
\end{center}
\end{figure}

This result gives a possible explanation for the enhanced 
residual resistivity inside the AFM phase under pressure in 
 CeAl$_2$\cite{CeAl2}, CeCu$_5$Au,\cite{CeCu5Au} and
 YbNi$_2$Ge$_2$\cite{YbNi2Ge2}, as schematically shown in
 Fig. \ref{fig-1}(a), because nonmagnetic impurity potential 
is enhanced by charge fluctuations,\cite{Nozieres} 
leading to the enhanced residual resistivity.
Interestingly, the N\'eel temperature $T_N$ and 
the Kondo temperature $T_K$ become comparable at the pressure where
the residual resistivity is enhanced. Recently, similar behavior is observed in 
CeRu$_2$(Ge$_x$Si$_{1-x}$)$_2$ at approximately $x\sim 0.8$.\cite{matsumoto}
It is also often the case that
 the residual resistivity is enhanced at QCP (see Fig. \ref{fig-1}(b)).
However, in these compounds,
 the enhanced residual resistivity is observed not at the QCP but inside
 the AFM phase. 
This phenomenon has not been 
well recognized or understood so far, and is 
interesting because the enhanced residual resistivity cannot be
explained by the critical AFM fluctuations near the QCP\cite{MiyakeNarikiyo,Kontani,exp}. 

As a preliminary step, we review how to estimate, 
at least qualitatively, the residual resistivity $\rho_0$ 
in heavy-fermion systems following the discussions in refs. 18 and 21; 
we focus on the nonmagnetic impurities, which
 are the dominant sources of the residual resistivity. 

First, note that the residual resistivity strongly 
depends on the effective impurity scattering strength that the 
quasiparticles are subjected to. One important point is that the quasiparticles 
mainly consist of f-electrons in heavy-fermion compounds. We will 
focus on such a situation hereafter. Thus, it is 
sufficient to take into account only the f-electron component of the 
quasiparticles. The nonmagnetic impurity scattering 
strength $u_{\bf q}$ is renormalized to $\tilde{u}_{\bf q}$ 
by the electron-electron 
interactions, as shown in Fig. \ref{fig-1}(c).
The residual resistivity is then given as $\rho_0\propto
N_F|\tilde{u}|^2$, with the renormalized 
density of states (DOS) at the Fermi level $N_F$.
From a detailed analysis of the renormalized $\tilde{u}_{\bf q}$ 
on the basis of the Fermi liquid theory in the 
periodic Anderson model\cite{Maebashi}, 
it is estimated that 
$\tilde{u}_{\bf q}\simeq -u_{\bf q}(dn_{\rm f}/d{\epsilon_{\rm f}})/N_F$
in heavy-fermion systems for ${\bf q}\to 0$, where $n_{\rm f}$ and $\epsilon_{\rm f}$ are the 
f-electron number per site and its energy level, respectively. 
Thus, in heavy-fermion systems, the residual resistivity is enhanced
when the f-electron charge susceptibility $\chi_{\rm f}\equiv-(dn_{\rm f}/d{\epsilon_{\rm f}})$  is enhanced. 


In this Letter, by using  Wilson's numerical renormalization group (NRG)
 method\cite{Wilson}, we will investigate the f-electron charge susceptibility
 $\chi_{\rm f}$ 
using 2IAM as a simplest example, which
 includes the competition between the Kondo effect and the intersite
 exchange interactions, and also the charge fluctuations of
 f-electrons. 
The model itself has been well understood,\cite{2impAnderson} together
 with a two-impurity
 Kondo model (2IKM),\cite{2impKondo,Affleck,Silva} in the last two
 decades. 
There are two stable fixed points. One is the Kondo-Yosida singlet fixed
 point, where f-electron spin degrees of freedom are 
screened by those of conduction electrons. The other is an intersite singlet fixed 
 point, where two f-electron spins form a spin-singlet state
 decoupled from conduction electrons. 
Between them, there is a NFL fixed point when the conduction 
electron bands exhibits particle-hole (PH) symmetry\cite{Affleck}. 

The Hamiltonian for 2IAM is given by
\begin{eqnarray}
H\!\!\!\!\!&=&\!\!\!\!\!\sum_{{\bf k}\sigma}\epsilon_{\bf k} c_{{\bf
 k}\sigma}^{\dagger}c_{{\bf k}\sigma}+\sum_{\sigma\alpha}\epsilon_{\rm
 f}
 f^{\dagger}_{\alpha\sigma}f_{\alpha\sigma}+U\sum_{\alpha}f^{\dagger}_{\alpha\uparrow}f_{\alpha\uparrow}f^{\dagger}_{\alpha\downarrow}f_{\alpha\downarrow}\nonumber\\
&&+\frac{v}{\sqrt{N_0}}\sum_{{\bf k}\sigma\alpha}[e^{{\rm i}{\bf k}{\bf x}_{\alpha}}f^{\dagger}_{\alpha\sigma}c_{{\bf k}\sigma}+{\rm h.c.}]
+J{\bf S}_1\cdot{\bf S}_2, \label{3}
\end{eqnarray}
where $f^{\dagger}_{\alpha\sigma}$ is an f-electron creation operator 
with spin $\sigma=\uparrow$ or $\downarrow$ and site index $\alpha=1$ or $2$, and 
${\bf S}_{\alpha}$ is a spin operator of an f-electron at postion ${\bf x}\alpha$. 
$c^{\dagger}_{{\bf k}\sigma}$ is a conduction electron creation operator 
with wave number $\bf k$ and spin $\sigma$. $N_0,\ \epsilon_{\bf k},\ U$ 
and $v$ are the number of sites, the kinetic energy of conduction
electrons, and the Coulomb repulsion and hybridization, respectively. 
$J$ is a phenomenological intersite exchange interaction between 
f-electrons and it is assumed to be antiferromagnetic, $J>0$. 
This is introduced to enlarge the parameter space of 2IAM. One of 
the physical interpretations of $J$ is, for example, the 
superexchange interaction due to the direct hopping processes of f-electrons.
For later purposes, we
define the distance between the two sites as $R\equiv |{\bf x}_{1}-{\bf x}_2|$. 
 We set $U=10D$ ($D$ being half the bandwidth
 of conduction electrons) and vary
$\epsilon_{\rm f}$, $v$, and $J$ to examine the variations of $\chi_{\rm f}$.

In the NRG calculation, we transform eq. (\ref{3}) into a form of 
semi-infinite one-dimensional chains, and diagonalize it iteratively.\cite{Wilson} 
In the two-impurity model, we have two conduction electron bands,
namely, even
and odd bands reflecting the mirror symmetry of the system\cite{2impKondo}. We use 
the logarithmic discretization parameter $\Lambda=3$ and the
low-energy 1000 states are kept at each NRG iteration.

First, we discuss the case of the PH-symmetric conduction electron bands. We use a constant DOS
 for both the even and odd conduction electrons in the following. 
 In the final part of this Letter, we will examine the
effects of PH asymmetry.

\begin{figure}[t]
\begin{center}
\includegraphics[scale =0.8]{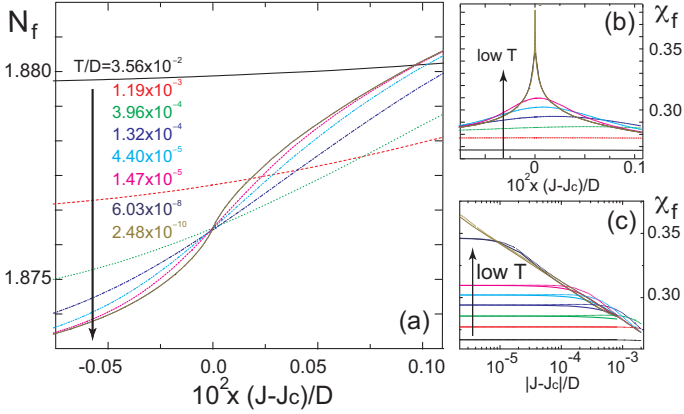}

\caption{(Color online) (a) Total f-electron number $N_{\rm f}$ vs $J-J_c$ for eight
 different temperatures for $v/D=0.18$ with PH-symmetric conduction electron bands. 
(b) f-electron charge susceptibility $\chi_{\rm f}(T)$ vs $J-J_c$. (c)
 $\chi_{\rm f}(T)$ for $J-J_c<0$ (dotted lines) and $J-J_c>0$ (full lines) on logarithmic scale. 
 }
\label{fig-2}
\end{center}
\end{figure}

Figure \ref{fig-2}(a) shows the total f-electron
number $N_{\rm f}\equiv\sum_{\sigma\alpha}\langle
f^{\dagger}_{\alpha\sigma}f_{\alpha\sigma}\rangle$ for $\epsilon_{\rm
f}=-0.8D$ and $v=0.18D$ as a function of $J-J_c$, where $\langle \cdots
\rangle$ indicates the thermal average and $J_c$ is the critical value
of $J$, $J_c=0.0018215D$, for this
parameter set. At high temperatures $T$, $N_{\rm f}$ gradually increases
 as $J$ increases, while the variation becomes steeper near $J=J_c$ as $T$ decreases.
Corresponding to this, $\chi_{\rm f}$ has a sharp peak at $J=J_c$ as a
function of $J$ at low temperatures, as shown in Figs. \ref{fig-2}(b)
and (c).

\begin{figure}[t]
\begin{center}
\includegraphics[scale =0.8]{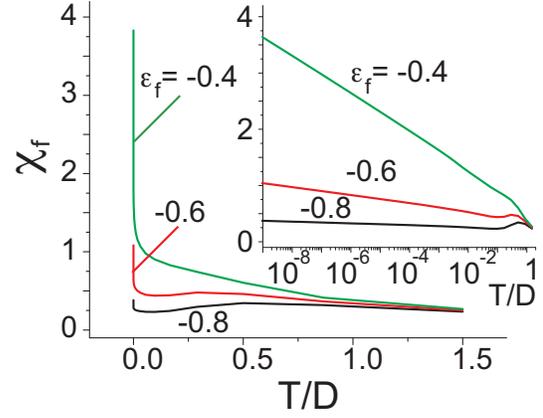}
\caption{(Color online) Temperature dependence of the f-electron charge
susceptibility $\chi_{\rm f}(T)$ at $J=J_c(\epsilon_{\rm f})$ for 
$\epsilon_{\rm f}/D=-0.4$, $-0.6$ and $-0.8$, and $v/D=0.18$ with 
PH-symmetric conduction electron bands. 
Inset: $\chi_{\rm f}(T)$ on logarithmic temperature scale.} 
\label{fig-3}
\end{center}
\end{figure}

Figure \ref{fig-2}(c) shows that the $J-J_c$ dependence of $\chi_{\rm f}(0)$ 
is logarithmic, {\it i.e.,} 
\begin{eqnarray}
\chi_{\rm f}(0)\propto -\log|J-J_c|\ \ \  {\rm for} \ |J-J_c|/J_c\ll1, \label{5}
\end{eqnarray}
at $T=0$.
As for the temperature dependence of $\chi_{\rm f}$, 
 $\chi_{\rm f}(T)$ just at the critical point 
is shown in Fig. \ref{fig-3} for three values of $\epsilon_{\rm f}$. 
Note that the charge fluctuation is 
enhanced as $|\epsilon_{\rm f}|$ decreases and  
$\chi_{\rm f}$ increases with decreasing $T$ and diverges at $T=0$. 
As shown in Fig. \ref{fig-3}, the 
temperature dependence of $\chi_{\rm f}(T)$ is also logarithmic:  
\begin{eqnarray}
\chi_{\rm f}(T)\propto -\log T\ \ \ {\rm for}\ J=J_c. \label{6}
\end{eqnarray}

Corresponding to the diverging f-electron charge susceptibility at $J=J_c$, 
the imaginary part of the dynamical f-electron charge susceptibility 
Im$\chi_{\rm f}(\omega)$ has a finite amplitude at the frequency 
$\omega=0$ and at $J=J_c$\cite{HatJMMM}, where a finite 
residual entropy $\log\sqrt{2}$ remains.\cite{Affleck2CK} 
Away from the critical point, Im$\chi_{\rm f}(\omega)$ 
is proportional to $\omega$ at low frequency, 
recovering the Fermi liquid behavior.

The above results suggest that there is a singularity in the charge sector 
through the formation of the Kondo-Yosida singlet between the f- and 
conduction electrons leading to a steep change in $N_{\rm f}$. 
The diverging $\chi_{\rm f}$ is related to the steep change in the 
phase shift between $\delta=0$ and $\pi/2$  in the case of  the PH-symmetric conduction 
electrons. In order to explain this, it is essential to note that 
the numbers of f-electrons 
are different between the two fixed points: the
f-electrons are more localized in the
inter-site spin-singlet fixed point than in the Kondo-Yosida singlet
fixed point, and the singular $|J-J_c|$ dependence in $\chi_{\rm f}$
 appears between
these two stable fixed points, {\it i.e.,} at the NFL fixed point.

On the basis of the boundary  
conformal field theoretical approach in 2IKM\cite{Affleck}, 
it is expected that the spin-singlet Cooper channel and the charge 
susceptibility of the conduction electrons diverge at the NFL fixed  
point. It is often the case that the charge susceptibility of conduction
electrons  
diverges at NFL fixed points in various 
impurity models, such as the multichannel\cite{Affleck2CK} and spin 3/2 multipolar\cite{Hat3half} Kondo models.

The origin of the logarithmic divergence $\chi_{\rm f}(T)\propto -\log T$ 
is related to the existence of boundary operators with the scaling  
dimension $\Delta=1/2$ near the critical point of 2IAM. Here, we assume that the operator contents at the
critical point of the 2IAM is the same as that in 2IKM,
 which is a reasonable assumption. 
The candidate is the charge vector field operator $\vec{ \phi}_c$ in Table VI in
ref. 25.
$\vec{ \phi}_c$ 
can be regarded as the f-electron charge operator and can 
couple with the first descendant of the total charge current 
operator ${\bf J}^c_{-1}$ around the NFL fixed point. 

Following the discussions in  
the multichannel Kondo model,\cite{Affleck2CK} 
we obtain the logarithmic temperature 
dependence of the charge susceptibility. 
We also obtain logarithmic divergence in the specific heat coefficient. 
As discussed by Johannesson {\it et al.,} in the context
of the two-channel Anderson model\cite{Bolech}, 
the important point is that the coefficient in front of the logarithm 
is expected to be scaled by 
 $1/D$ rather than by $1/T_K$, leading to a small coupling constant. 
Then, we can safely neglect it in the specific heat calculation. 
However, in the calculation of the charge susceptibility, we cannot
neglect it, since there is no contribution of other operators here. 
This small coefficient is
the reason why we find moderate divergence in the f-electron charge
susceptibility in Figs. \ref{fig-2} and \ref{fig-3}. 

In order to discuss a more realistic situation, we introduce 
PH asymmetry to the conduction electrons below. 
The importance of the PH symmetry in 2IKM 
and 2IAM has been studied in detail\cite{Affleck,2impAnderson}. 
Indeed, the model 
lacks the critical point present in the PH-symmetric 
case. The two different fixed points with phase shifts 
$\delta=0$ and $\pi/2$ can be connected by a smooth crossover. Although 
it is not a transition but a crossover, the nonmonotonic variation in  
$\chi_{\rm f}$ is
expected owing to the ``hidden''  NFL fixed point. 
In the numerical calculations below, we set 
$\epsilon_{\bf k}\simeq v_F(k-k_F)$, 
 $k_F R=\pi$ and $D=v_Fk_F/\pi$, 
where $v_F$ and $k_F$ are the Fermi velocity and
 momentum, respectively. For these parameters, an intersite exchange 
interaction is induced, which is antiferromagnetic\cite{Silva}. Thus, we set $J=0$.
The density of states $\rho_{\pm}$ for the even
 $(+)$ and odd $(-)$
 bands of conduction electrons are given as
$\rho_{\pm}=A_{\pm} [ 1\pm (\sin kR)/kR ]$,
where $A_{\pm}$ is the normalization constant.\cite{Cox}

\begin{figure}[t!]
\begin{center}
\includegraphics[scale =0.8]{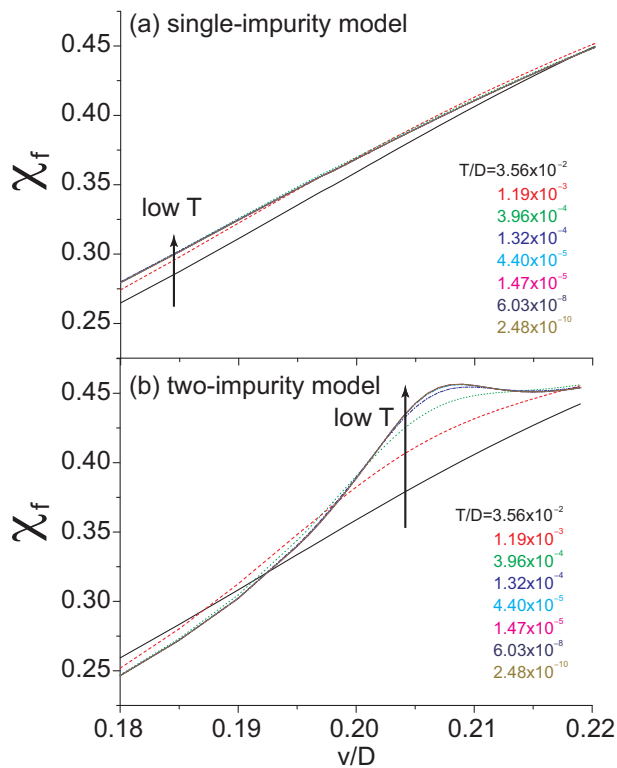}
\caption{(Color online) f-Electron charge susceptibility $\chi_{\rm f}(T)$ vs $v$ for
 $\epsilon_{\rm f}/D=-0.8$, $J=0$. (a) 
 Single impurity model with the constant DOS, and (b)
 two-impurity model without PH symmetry. For (a), the value is
 multiplied by the factor 2.}
\label{fig-5}
\end{center}
\end{figure}

Figure {\ref{fig-5}} shows the $v$ dependence of $\chi_{\rm f}$ for eight
different temperatures for (a) the single-impurity Anderson model with a 
constant DOS and 
(b) 2IAM. As expected, $\chi_{\rm f}(T)$ for the 
single-impurity model shows a monotonic increase as 
$v$ increases in (a). This is natural since as $v$ increases, the
charge fluctuations of f-electrons are enhanced.
 In the case of the 2IAM, as shown
in (b), $\chi_{\rm f}(T)$ develops a peak structure at 
low temperatures at approximately around 
$v\sim 0.205D$ in addition to the monotonic increase as $v$
increases. Around $v=0.205D$, the Kondo-Yosida singlet starts to form 
as $v$ increases, while for $v\simle 0.205D$ two f-electrons form a spin
singlet owing to the induced antiferromagnetic interaction.

As discussed above, 
we have demonstrated that $\chi_{\rm f}$ is enhanced near the
critical point and the crossover region between the Kondo-Yosida singlet and 
the intersite spin-singlet states. By combining this result with those in refs. 18 and 21, it is expected that the residual resistivity
is enhanced owing to the Kondo-Yosida singlet formation. 
Although our analysis is based on a 
two-impurity problem, we expect that the same phenomenon exists in the 
 periodic lattice models for heavy-fermion systems. 
More elaborate calculations are necessary 
to examine the present scenario for the enhanced residual resistivity in 
heavy-fermion systems. 

As for the Fermi surfaces in the Anderson lattice model, 
the occurrence of the Kondo-Yosida singlet formation strongly 
affects the topology of the Fermi surfaces inside the AFM phase.
From this point, it is expected that the residual resistivity anomaly
 will have a close
relation with the changes in the Fermi surface. 
In real materials, as shown in Fig. \ref{fig-1}(a), the 
system is inside the AFM phase at low temperatures when the residual resistivity
shows enhancement as a function of pressure.
In the AFM phase, 
the topology of Fermi surfaces in the Anderson lattice model
is not yet fully understood and it is also nontrivial to regard it to 
be the same as that in the Kondo lattice model. A recent 
variational Monte Carlo study shows 
 the existence of 
two different phases in the AFM phase\cite{Watanabe}.  
Further theoretical studies are necessary to clarify the 
relationship between the Kondo-Yosida singlet formation and 
the enhanced residual resistivity.

In summary, we have discussed the f-electron charge susceptibility
$\chi_{\rm f}(T)$ near the critical point of the two-impurity Anderson
model. We have found that $\chi_{\rm f}(T)$ diverges at the critical
point of this model with particle-hole-symmetric conduction electrons.
Even when the conduction electron bands are not 
particle-hole-symmetric and thus the critical point smears out, 
$\chi_{\rm f}(T)$ is still much more  
enhanced than that in the single-impurity case around 
the crossover regime between the Kondo-Yosida singlet and intersite
spin-singlet fixed points. These results shed light on the effects
of intersite correlation on heavy-fermion 
formation and the enhanced residual resistivity 
in Ce- and Yb-based heavy-fermion compounds.
\section*{Acknowledgments}
K. H. acknowledge useful discussions with  C. M. Varma and L. Zhu. 
K. H. also acknowledge hospitality from the University of California,
Riverside, while part of this work was done. This work was supported by a Grant-in-Aid for Scientific
Research (No. 20740189) from the Japan Society for the Promotion of Science,
and by a Grant-in-Aid for Scientific Research in Innovative Areas ``Heavy
Electrons'' (No. 20102008) from the Ministry of Education, Culture,
Sports, Science and Technology, Japan.


\begin{thebibliography}{99}
\bibitem{exp} In the case shown in Fig. \ref{fig-1}(b), there also
	remains the possibility that the charge fluctuations associated with
	the Kondo-Yosida singlet formation at the QCP enhances residual resistivity.
\end{thebibliography}
\end{document}